\documentclass[]{article}
\usepackage{lmodern}
\usepackage{amssymb,amsmath}
\usepackage{ifxetex,ifluatex}
\usepackage{fixltx2e} % provides \textsubscript
\ifnum 0\ifxetex 1\fi\ifluatex 1\fi=0 % if pdftex
  \usepackage[T1]{fontenc}
  \usepackage[utf8]{inputenc}
\else % if luatex or xelatex
  \ifxetex
    \usepackage{mathspec}
    \usepackage{xltxtra,xunicode}
  \else
    \usepackage{fontspec}
  \fi
  \defaultfontfeatures{Mapping=tex-text,Scale=MatchLowercase}
  
\fi
\IfFileExists{upquote.sty}{\usepackage{upquote}}{}
\IfFileExists{microtype.sty}{%
\usepackage{microtype}
\UseMicrotypeSet[protrusion]{basicmath} % disable protrusion for tt fonts
}{}
\usepackage[margin=1in]{geometry}
\ifxetex
  \usepackage[setpagesize=false, % page size defined by xetex
              unicode=false, % unicode breaks when used with xetex
              xetex]{hyperref}
\else
  \usepackage[unicode=true]{hyperref}
\fi
\hypersetup{breaklinks=true,
            bookmarks=true,
            pdfauthor={},
            pdftitle={},
            colorlinks=true,
            citecolor=blue,
            urlcolor=blue,
            linkcolor=magenta,
            pdfborder={0 0 0}}
\usepackage{fancyhdr}
\usepackage{longtable,booktabs}
\usepackage{graphicx,grffile}
\makeatletter
\def\maxwidth{\ifdim\Gin@nat@width>\linewidth\linewidth\else\Gin@nat@width\fi}
\def\maxheight{\ifdim\Gin@nat@height>\textheight\textheight\else\Gin@nat@height\fi}
\makeatother
\setkeys{Gin}{width=\maxwidth,height=\maxheight,keepaspectratio}
\date{}

\ifx\paragraph\undefined\else
\let\oldparagraph\paragraph
\renewcommand{\paragraph}[1]{\oldparagraph{#1}\mbox{}}
\fi
\ifx\subparagraph\undefined\else
\let\oldsubparagraph\subparagraph
\renewcommand{\subparagraph}[1]{\oldsubparagraph{#1}\mbox{}}
\fi

\begin{document}

\section{Epiphany-V: A 1024 processor 64-bit RISC
System-On-Chip}\label{epiphany-v-a-1024-processor-64-bit-risc-system-on-chip}

By Andreas Olofsson\\Adapteva Inc, Lexington, MA,
USA\\andreas@adapteva.com

\subsection{Abstract}\label{abstract}

This paper describes the design of a 1024-core processor chip in 16nm
FinFet technology. The chip (``Epiphany-V'') contains an array of 1024
64-bit RISC processors, 64MB of on-chip SRAM, three 136-bit wide mesh
Networks-On-Chip, and 1024 programmable IO pins. The chip has taped out
and is being manufactured by TSMC.

This research was developed with funding from the Defense Advanced
Research Projects Agency (DARPA). The views, opinions and/or findings
expressed are those of the author and should not be interpreted as
representing the official views or policies of the Department of Defense
or the U.S. Government.

\textbf{Keywords:} RISC, Network-on-Chip (NoC), HPC, parallel,
many-core, 16nm FinFET

\subsection{I. Introduction}\label{i.-introduction}

Applications like deep learning, self-driving cars, autonomous drones,
and cognitive radio need an order of magnitude boost in processing
efficiency to unlock their true potentials. The primary goal for this
project is to build a parallel processor with 1024 RISC cores
demonstrating a processing energy efficiency of 75 GFLOPS/Watt. A
secondary goal for this project is to demonstrate a 100x reduction in
chip design costs for advanced node ASICs. Significant energy savings of
10-100x can be achieved through extreme silicon customization, but
customization is not financially viable if chip design costs are
prohibitive. The general consensus is that it costs anywhere from \$20M
to \$1B to design a leading edge System-On-Chip platform.{[}1-4{]}

\subsection{II. History}\label{ii.-history}

The System-On-Chip described in this paper is the 5th generation of the
Epiphany parallel processor architecture invented by Andreas Olofsson in
2008.{[}5{]} The Epiphany architecture was created to address energy
efficiency and peak performance limitations in real time communication
and image processing applications.

The first Epiphany product was a 16-core 65nm System-On-Chip
(``Epiphany-III'') released in May 2011. The chip worked beyond
expectations and is still being produced today.{[}6{]}

The second Epiphany product was a 28nm 64-core SOC (``Epiphany-IV'')
completed in the summer of 2011.{[}7{]} The Epiphany-IV chips
demonstrated 70 GFLOPS/Watt processing efficiency at the core supply
level and was the most energy-efficient processor available at that
time. The chip was sampled to a number of customers and partners, but
was not produced in volume due to lack of funding. At that time,
Adapteva also created a physical implementation of a 1024 core 32-bit
RISC processor array, but it was never taped out due to funding
constraints.

In 2012 Adapteva launched an open source \$99 Epiphany-III based single
board computer on Kickstarter called Parallella.{[}8{]} The goal of the
project was to democratize access to parallel computing for researchers
and programming enthusiasts. The project was highly successful and
raised close to \$1M on Kickstarter. To date the Parallella computer has
shipped to over 10,000 customers and has generated over 100 technical
publications.{[}9{]}

For a complete description of the Epiphany processor history and design
decisions, please refer to the paper ``Kickstarting high-performance
energy-efficient manycore architectures with Epiphany''.{[}10{]}

\subsection{III. Architecture}\label{iii.-architecture}

\subsubsection{III.A Overview}\label{iii.a-overview}

The Epiphany architecture is a distributed shared memory architecture
comprised of an array of RISC processors communicating via a low-latency
mesh Network-on-Chip. Each node in the processor array is a complete
RISC processor capable of running an operating system (``MIMD'').
Epiphany uses a flat cache-less memory model, in which all distributed
memory is readable and writable by all processors in the system.

The Epiphany-V introduces a number of new capabilities compared to
previous Epiphany products, including 64-bit memory addressing, 64-bit
floating point operations, 2X the memory per processor, and custom ISAs
for deep learning, communication, and cryptography. The following figure
shows a high level diagram of the Epiphany-V implementation.

\begin{figure}[htbp]
\centering
\includegraphics{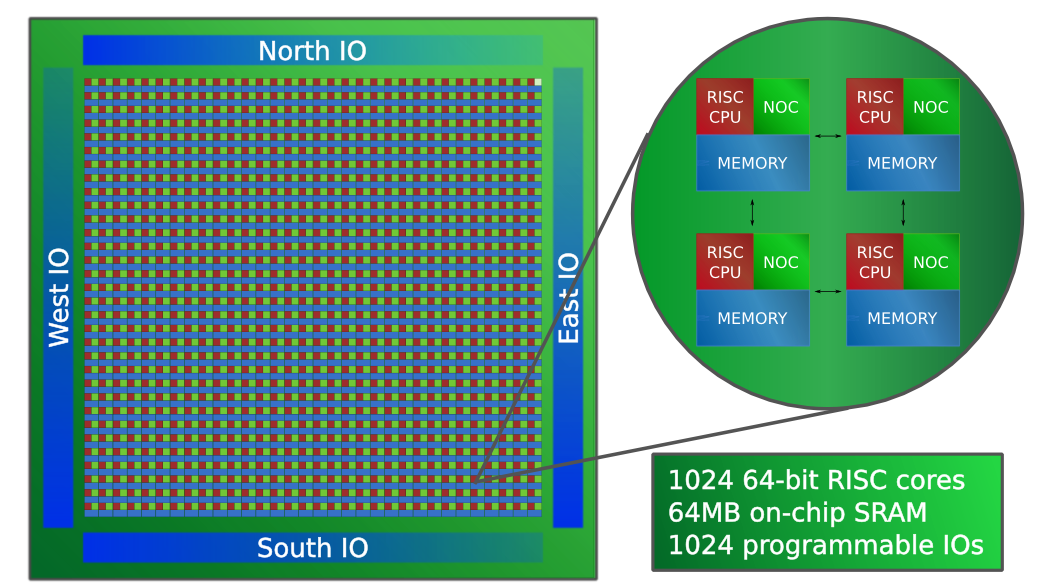}
\caption{Epiphany-V Overview}
\end{figure}

Summary of Epiphany-V features:

\begin{itemize}
\itemsep1pt\parskip0pt\parsep0pt
\item
  1024 64-bit RISC processors
\item
  64-bit memory architecture
\item
  64/32-bit IEEE floating point support
\item
  64MB of distributed on-chip memory
\item
  1024 programmable I/O signals
\item
  Three 136-bit wide 2D mesh NOCs
\item
  2052 Independent Power Domains
\item
  Support for up to 1 billion shared memory processors
\item
  Binary compatibility with Epiphany III/IV chips
\item
  Custom ISA extensions for deep learning, communication, and
  cryptography
\end{itemize}

As in previous Epiphany versions, multiple chips can be connected
together at the board and system level using point to point links.
Epiphany-V has 128 point-to-point I/O links for chip to chip
communication. In aggregate, the Epiphany 64-bit architecture supports
systems with up to 1 Billion cores and 1 Petabyte (10\^{}15) of total
memory.

\begin{figure}[htbp]
\centering
\includegraphics{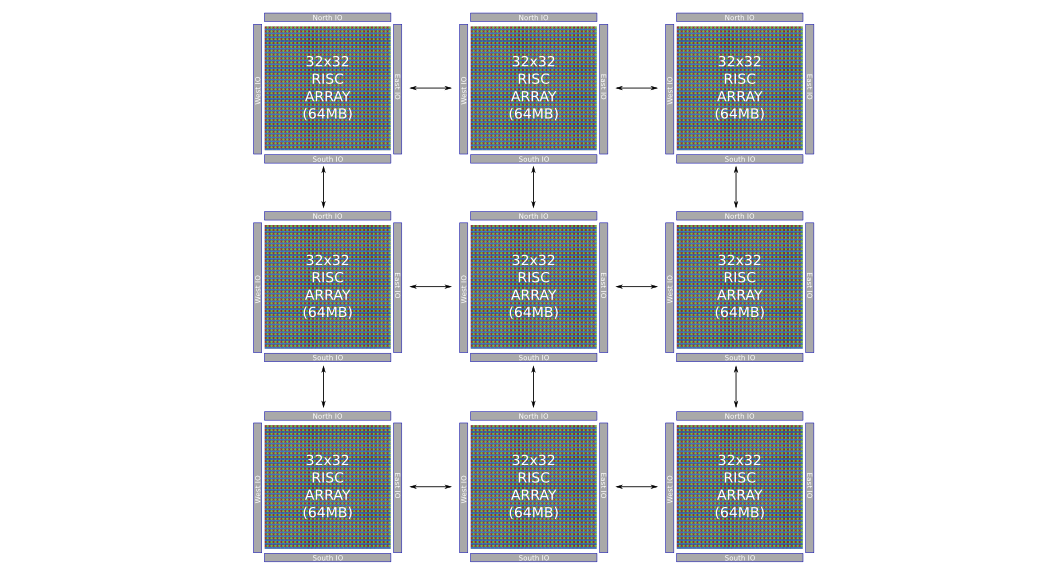}
\caption{Multichip configuration}
\end{figure}

The following sections describe the Epiphany architecture. For complete
details, please refer to the online architecture reference
manual.{[}11{]}

\subsubsection{III.B Memory
Architecture}\label{iii.b-memory-architecture}

The Epiphany 64-bit memory map is split into 1 Billion 1MB memory
regions, with 30 bits dedicated to x,y,z addressing. The complete
Epiphany memory map is flat, distributed, and shared by all processors
in the system. Each individual memory region can be used by a single
processor or aggregated as part of a shared memory pool. The Epiphany
architecture uses multi-banked software-managed scratch-pad memory at
each processor node. On every clock cycle, a processor node can:

\begin{itemize}
\itemsep1pt\parskip0pt\parsep0pt
\item
  Fetch 8 bytes of instructions
\item
  Load/store 8 bytes of data
\item
  Receive 8 bytes from another processor in the system
\item
  Send 8 bytes to another processor in the system
\end{itemize}

The Epiphany architecture uses strong memory ordering for local
load/stores and weak memory ordering remote transfers.

\newpage

\begin{longtable}[c]{@{}lll@{}}
\toprule\addlinespace
Transfer \#1 & Transfer \#2 & Deterministic
\\\addlinespace
\midrule\endhead
Read Core A & Read Core A & Yes
\\\addlinespace
Read Core A & Read Core B & Yes
\\\addlinespace
Read Core A & Write Core A & Yes
\\\addlinespace
Read Core A & Write Core B & Yes
\\\addlinespace
Write Core A & Write Core A & Yes
\\\addlinespace
Write Core A & Write Core B & No
\\\addlinespace
Write Core A & Read Core A & No
\\\addlinespace
Write Core A & Read Core B & No
\\\addlinespace
\bottomrule
\addlinespace
\caption{Epiphany Remote Transfer Memory Order}
\end{longtable}

\subsubsection{III.C Network-On-Chip}\label{iii.c-network-on-chip}

The Epiphany-V mesh Network-on-Chip (``emesh'') consists of three
independent 136-bit wide mesh networks. Each one of the three NOCs serve
different purposes:

\begin{itemize}
\itemsep1pt\parskip0pt\parsep0pt
\item
  rmesh: Read request packets
\item
  cmesh: On-chip write packets
\item
  xmesh: Off-chip write packets
\end{itemize}

Epiphany NOC packets are 136 bits wide and transferred between
neighboring nodes in one and a half clock cycles. Packets consist of 64
bits of data, 64 bits of address, and 8 bits of control. Read requests
puts a second 64-bit address in place of the data to indicate
destination address for the returned read data.

Network-On-Chip routing follows a few simple, static rules. At every
hop, the router compares its own coordinate address with the packet's
destination address. If the column addresses are not equal, the packet
gets immediately routed to the south or north; otherwise, if the row
addresses are not equal, the packet gets routed to the east or west;
otherwise the packet gets routed into the hub node.

Each routing node consists of a round robin five direction arbiter and a
single stage FIFO. Single cycle transaction push-back enables network
stalling without packet loss.

\subsubsection{III.D Processor}\label{iii.d-processor}

The Epiphany includes is an in-order dual-issue RISC processor with the
following key features:

\begin{itemize}
\itemsep1pt\parskip0pt\parsep0pt
\item
  Compressed 16/32-bit ISA
\item
  IEEE-754 compatible floating-point instruction set (FPU)
\item
  Integer arithmetic logic instruction set (IALU)
\item
  Byte addressable load/store instructions with support for 64-bit
  single cycle access
\item
  64-word 6 read/3-write port register file
\end{itemize}

Several new processor features have been introduced in the Epiphany-V
chip:

\begin{itemize}
\itemsep1pt\parskip0pt\parsep0pt
\item
  64/32-bit addressing
\item
  64-bit integer instructions
\item
  64-bit IEEE floating point support
\item
  SIMD 32-bit IEEE floating point support
\item
  Expanded shared memory support for up to 1 Billion cores
\item
  Custom ISA extensions for deep learning, communication, and
  cryptography
\end{itemize}

\subsubsection{III.E I/O}\label{iii.e-io}

The Epiphany-V has a total of 1024 programmable I/O pins and 16 control
input pins. The programmable I/O is configured through 32 independent IO
modules ``io-slices'' on each side of the chip (north, east, west,
south). All io-slices can be independently configured as fast
point-to-point links or as GPIO pins.

When the IO modules are configured as links, epiphany memory
transactions are transferred across the IO links automatically,
effectively extending the on-chip 2D mesh network to other chips. The
glueless memory transfer point-to-point I/O links combined with 64-bit
addressability enables construction of shared memory systems with up to
1 Billion Epiphany processors.

\subsection{IV. Performance}\label{iv.-performance}

The following table illustrates aggregate frequency independent
performance metrics for the Epiphany-V chip. Actual Epiphany-V
performance numbers will be disclosed once silicon chips have been
tested and characterized.

\begin{longtable}[c]{@{}ll@{}}
\toprule\addlinespace
Metric & Value
\\\addlinespace
\midrule\endhead
64-bit FLOPS & 2048 / clock cycle
\\\addlinespace
32-bit FLOPS & 4096 / clock cycle
\\\addlinespace
Aggregate Memory Bandwidth & 32,768 Bytes / clock cycle
\\\addlinespace
NOC Bisection Bandwidth & 1536 Bytes / clock cycle
\\\addlinespace
IO Bandwidth & 192 Bytes / IO clock cycle
\\\addlinespace
\bottomrule
\addlinespace
\caption{Epiphany-V Processor Performance}
\end{longtable}

\subsection{V. Programming Model}\label{v.-programming-model}

Each Epiphany RISC processor is programmable in ANSI-C/C++ using a
standard open source GNU tool chain based on GCC-5 and GDB-7.10.

Mapping complicated algorithm to massively parallel hardware
architectures is considered a non trivial problem. To ease the challenge
of parallel programming, the Parallella community has created a number
of high quality parallel programming frameworks for the Epiphany.

A 1024 core functional simulator has been developed for Epiphany-V to
simplify porting legacy software from Epiphany-III. Several examples,
including matrix-matrix multiplication has been ported to Epiphany-V and
run on the new simulator with minimal engineering effort.

\newpage

\begin{longtable}[c]{@{}lll@{}}
\toprule\addlinespace
Framework & Author & Reference
\\\addlinespace
\midrule\endhead
OpenMP & University of Ioannina & {[}12{]}
\\\addlinespace
MPI & BDT/ARL & {[}13{]}
\\\addlinespace
OpenSHMEM & ARL & {[}14{]}
\\\addlinespace
OpenCL & BDT & {[}15{]}
\\\addlinespace
Erlang & Uppsala University & {[}16{]}
\\\addlinespace
Bulk Synchronous Parallel (BSP) & Coduin & {[}17{]}
\\\addlinespace
Epython & Nick Brown & {[}18{]}
\\\addlinespace
PAL & Adapteva/community & {[}19{]}
\\\addlinespace
\bottomrule
\addlinespace
\caption{Supported Epiphany Programming Frameworks}
\end{longtable}

\subsection{VI. Chip Implementation}\label{vi.-chip-implementation}

\subsubsection{VI.A Physical Design
Details}\label{vi.a-physical-design-details}

Given the complexity of advanced technology chip design, it is not
advisable to change too many design parameters at one time. Intel has
demonstrated commercial success over the last decade using the
conservative ``Tick-Tock'' model. In contrast, the ambitious Epiphany-V
chip described in this paper involved a new 64-bit architecture,
rewriting 95\% of the Epiphany RTL code, new EDA tools, new IP, and a
new processor node!

\begin{longtable}[c]{@{}ll@{}}
\toprule\addlinespace
Parameter & Value
\\\addlinespace
\midrule\endhead
Technology & TSMC 16nm FF+
\\\addlinespace
Metal Layers & 9
\\\addlinespace
VTH Types & 3
\\\addlinespace
Die Area & 117.44 mm\^{}2
\\\addlinespace
Transistors & 4.56B
\\\addlinespace
Flip-Chip Bumps & 3460
\\\addlinespace
IO Signal Pins & 1040
\\\addlinespace
Clock Domains & 1152
\\\addlinespace
Voltage Domains & 2052
\\\addlinespace
\bottomrule
\addlinespace
\caption{Epiphany-V Physical Specifications}
\end{longtable}

\subsubsection{VI.B Design Methodology}\label{vi.b-design-methodology}

Since 2008, the Epiphany implementation methodology has involved abutted
tiled layout, distributed clocking, and point-to-point communication.
The following design principles have been strictly followed at all
stages of the architecture and chip development:

\begin{itemize}
\itemsep1pt\parskip0pt\parsep0pt
\item
  Symmetry
\item
  Modularity
\item
  Scalability
\item
  Simplicity
\end{itemize}

The Epiphany-V required significant advances to accommodate large array
size and advanced process technology node. Novel circuit topologies were
created to solve critical issues in the areas of clocking, reset, power
grids, synchronization, fault tolerance, and standby power.

\subsubsection{VI.C Chip Layout}\label{vi.c-chip-layout}

This section includes the silicon area breakdown of the Epiphany-V and
layout screen-shots demonstrating the scalable implementation
methodology. The exact chip size including the chip guard ring is
15076.550um by 7790.480um.

\begin{longtable}[c]{@{}lcc@{}}
\toprule\addlinespace
Function & Value (mm\^{}2) & Share of Total Die Area
\\\addlinespace
\midrule\endhead
SRAM & 62.4 & 53.3\%
\\\addlinespace
Register File & 15.1 & 12.9\%
\\\addlinespace
FPU & 11.8 & 10.1\%
\\\addlinespace
NOC & 12.1 & 10.3\%
\\\addlinespace
IO Logic & 6.5 & 5.6\%
\\\addlinespace
``Other'' Core Stuff & 5.1 & 4.4\%
\\\addlinespace
IO Pads & 3.9 & 3.3\%
\\\addlinespace
Always on Logic & 0.66 & 0.6\%
\\\addlinespace
\bottomrule
\addlinespace
\caption{Epiphany-V Area Breakdown}
\end{longtable}

Figure 3 shows the OD and poly mask layers of the Epiphany chip. The
most striking feature of the plot is the level of symmetry. The strong
deviation from a ``square die'' was due to the aspect ratio of the SRAM
macros and 16nm poly vertical alignment restriction.

Figure 4 shows the flip-chip bump layout. The symmetry of the Epiphany-V
architecture made flip-chip bump planning trivial. The chip contains a
total of 3460 flip-chip bumps at a minimum C4 bump pitch of 150um.
Signal bumps are placed around the periphery of the die while core power
and ground bumps are placed in the center area.

Figure 5 shows aspects of the abutted layout flow. The Epiphany-V top
level layout integration is done 100\% through connection by abutment.
Attempts at implementing other integration methods were unsuccessful due
to the size of the chip and server memory constraints.

Figure 6 shows the tile layout. Routing convergence at 16nm proved to be
significantly more challenging than previous efforts at 28nm and 65nm.
The figure illustrates the final optimized processor tile layout after
iterating through many non-optimal configurations. Highlighted is the
logic for the NOC (green), FPU (blue), register file (orange), 4 memory
banks (2 on each side), and a small always on logic area (square blue).

Figure 7 shows qualitative IR drop analysis for a power gated power
rail. Power delivery to the core was implemented using a dense M8/M9
grid and sparse lower level metal grids. All tiles with the exception of
a small number of blocked peripheral tiles have individual flip-chip
power and ground bumps placed directly above the tile.

\begin{figure}[htbp]
\centering
\includegraphics{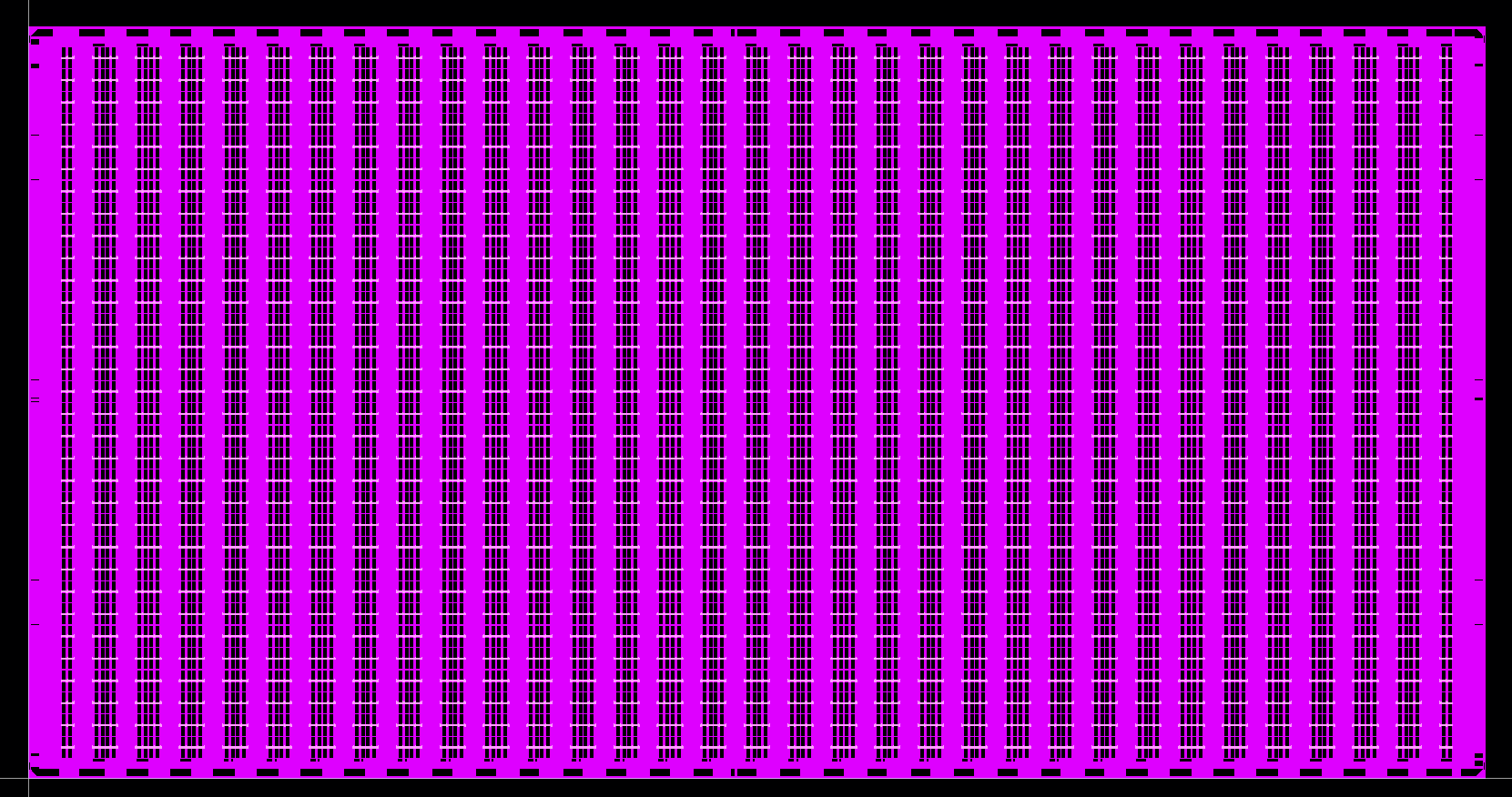}
\caption{Full Chip Layout (poly/od layers)}
\end{figure}

\begin{figure}[htbp]
\centering
\includegraphics{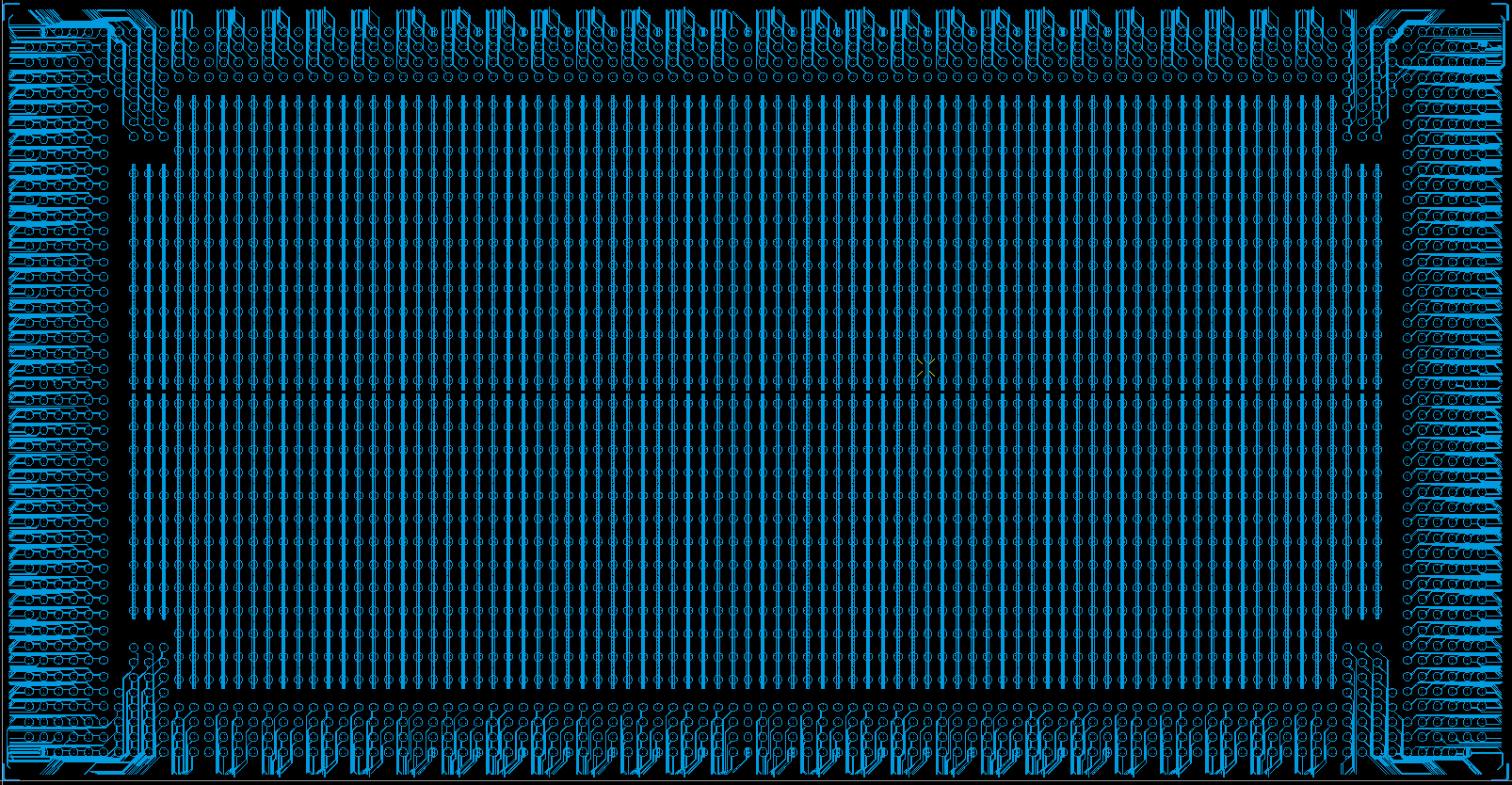}
\caption{Flip-Chip Bumps}
\end{figure}

\begin{figure}[htbp]
\centering
\includegraphics{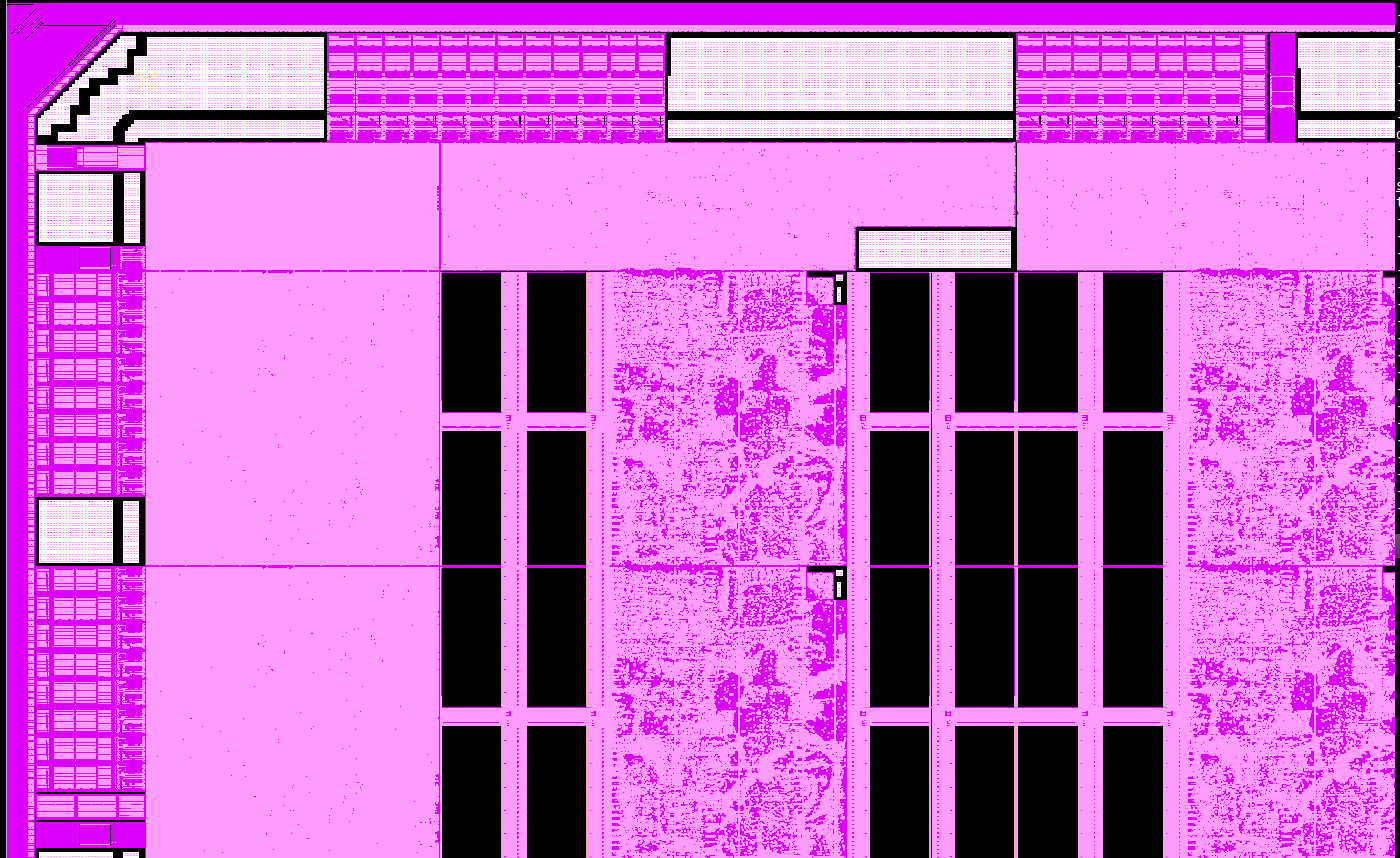}
\caption{Upper Left Chip Corner}
\end{figure}

\begin{figure}[htbp]
\centering
\includegraphics{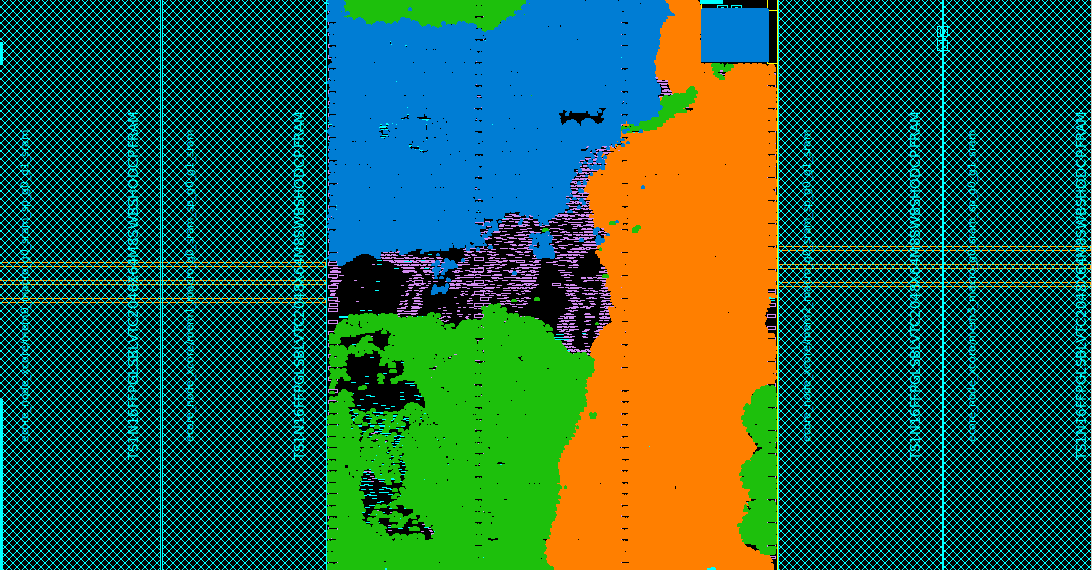}
\caption{Processor Node Layout}
\end{figure}

\begin{figure}[htbp]
\centering
\includegraphics{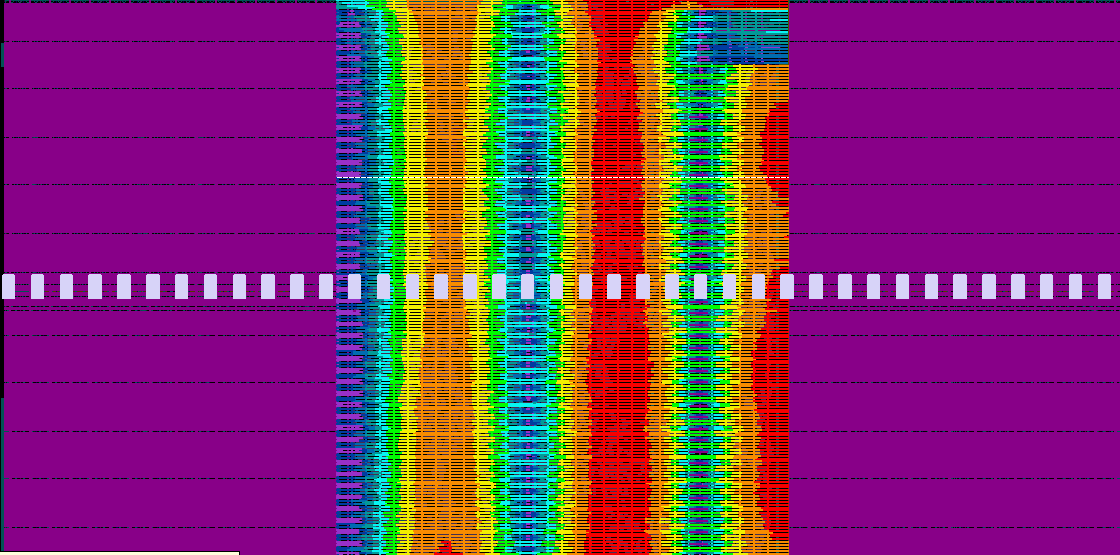}
\caption{Processor Node Power Grid Analysis}
\end{figure}

\newpage

\subsubsection{VI.D Chip Source Code}\label{vi.d-chip-source-code}

The Epiphany-V was designed using a completely automated flow to
translate Verilog RTL source code to a tapeout ready GDS, demonstrating
the feasibility of a 16nm ``silicon compiler''. The amount of open
source code in the chip implementation flow should be close to 100\% but
we were forbidden by our EDA vendor to release the code. All
non-proprietary RTL code was developed and released continuously
throughout the project as part of the ``OH!'' open source hardware
library.{[}20{]} The Epiphany-V likely represents the first example of a
commercial project using a transparent development model pre-tapeout.

\begin{longtable}[c]{@{}lccc@{}}
\toprule\addlinespace
Code & Language & LOC & Open Source \%
\\\addlinespace
\midrule\endhead
RTL & Verilog & 61K & 18\%
\\\addlinespace
Chip Implementation Code & TCL & 9K & \textless{}10\%
\\\addlinespace
Design Verification & C++ & 9K & 90\%
\\\addlinespace
\bottomrule
\addlinespace
\caption{Chip Code Base}
\end{longtable}

\subsubsection{VI.E Design Run Times}\label{vi.e-design-run-times}

Epiphany-V RTL to GDS run times were constrained by EDA license costs
and would take between 18 and 30 hrs. With an unlimited number of DRC,
synthesis, and place and route licenses and adequate hardware, the RTL
to GDS turnaround time would be less than 8hrs. All work was done on a
single Dell PowerEdge T610 purchased in 2010 with a quad-core Intel Xeon
5500 processor and 32GB of DDR3 memory.

\newpage

\begin{longtable}[c]{@{}lllll@{}}
\toprule\addlinespace
Step & Block A & Block B & Block C & Chip Level
\\\addlinespace
\midrule\endhead
Synthesis & 0.05 (x4) & 0.13 (x4) & 0.4 & 0
\\\addlinespace
PNR & 0.28 (x4) & 1.66 (x4) & 3.66 & 1
\\\addlinespace
Fill & 0.03 (x4) & 0.03 (x4) & 0.066 & 5
\\\addlinespace
DRC & 0 & 0 & 0 & 11
\\\addlinespace
\textbf{Total} & 1.46 hrs & 7.3 hrs & 4.13 hrs & 17 hrs
\\\addlinespace
\bottomrule
\addlinespace
\caption{Chip Generator Run Times}
\end{longtable}

\subsubsection{VI.F Chip Design Costs}\label{vi.f-chip-design-costs}

One of the goals of this research was to improve chip design cost
efficiency by 100x. Adapteva has previously shown the ability to design
chips at a fraction of the status-quo, but a 1024 core design at 16nm
would stretch that capability to the limit.{[}21-22{]} A major
contributing factor for SOC design cost explosion is the number of
complexity related stall cycles encountered by large design teams and
the enormous cost of each stall cycle. A design team of 100 US engineers
carries an effective cost of over \$50,000 per day, regardless of design
productivity.

Due to the scale of the challenges faced by the Epiphany-V related to
process migration, architecture co-development, RTL rewrite, and EDA
flow rampup, the project was in a constant state of flux, causing stall
cycles on a daily basis. The project was kicked off September 9th, 2015
with a design team consisting of Andreas Olofsson, Ola Jeppsson, and two
part time contractors. From January 2016 through tapeout in the summer
of 2016, design stall cycles forced Andreas Olofsson to complete the
project alone to stay within the fixed-cost DARPA budget. The tapeout of
a 1024-core 16nm processor in less than one year with a skeleton team
demonstrate it's possible to design advanced ASICs at 1/100th the cost
of the status quo.

\begin{longtable}[c]{@{}lll@{}}
\toprule\addlinespace
Designer & Responsibility & Effort (hrs)
\\\addlinespace
\midrule\endhead
Contractor A & Floating Point Unit & 200
\\\addlinespace
Contractor B & Design Verification Engine & 200
\\\addlinespace
Contractor C & EDA Tool support & 112
\\\addlinespace
Ola Jeppsson & Simulator/SDK & 500
\\\addlinespace
Andreas Olofsson & Everything else\ldots{} & 4100
\\\addlinespace
\bottomrule
\addlinespace
\caption{Chip Design Engineering Hours}
\end{longtable}

\begin{longtable}[c]{@{}ll@{}}
\toprule\addlinespace
Task & Wall Time
\\\addlinespace
\midrule\endhead
Architecture & 1 months
\\\addlinespace
RTL & 3 months
\\\addlinespace
IP integration & 1 months
\\\addlinespace
EDA methodology & 3 months
\\\addlinespace
Implementation & 2 months
\\\addlinespace
\bottomrule
\addlinespace
\caption{Chip Design Wall Times}
\end{longtable}

\begin{longtable}[c]{@{}ll@{}}
\toprule\addlinespace
Epiphany-V World Record/First & Mark
\\\addlinespace
\midrule\endhead
Chip with largest \# of General Purpose Processors & 1024
\\\addlinespace
Highest Density HPC Chip & 38M transistors/mm\^{}2
\\\addlinespace
Most efficient chip design team & 900K transistors/hour
\\\addlinespace
Most efficient RTL to GDS Chip Design flow & 150M transistors/hour
\\\addlinespace
Largest chip designed by one full time designer & 4.5B
\\\addlinespace
\bottomrule
\addlinespace
\caption{Epiphany-V Design Efficiency Benchmarks}
\end{longtable}

\subsection{VII. Competitive Data}\label{vii.-competitive-data}

The following tables compare the Epiphany-V chip and a selection of
modern parallel processor chips. The data shows Epiphany-V has an 80x
processor density advantage, and a 3.6x-15.8x memory density advantage
compared to the state of the art in parallel processors.

\begin{longtable}[c]{@{}lllllllll@{}}
\toprule\addlinespace
Chip & Company & Nodes & FLOPS & Area & Transistors & Power & Process &
Ref
\\\addlinespace
\midrule\endhead
P100 & Nvidia & 56 & 4.7T & 610 & 15.3B & 250W & 16FF+ & {[}23{]}
\\\addlinespace
KNL & Intel & 72 & 3.6T & 683 & 7.1B & 245W & 14nm & {[}24{]}
\\\addlinespace
Broadwell & Intel & 24 & 1.3T & 456 & 7.2B & 145W & 14nm & {[}25{]}
\\\addlinespace
Kilocore & UC-Davis & 1000 & N/A & 64 & 0.6B & 39W & 32nm & {[}26{]}
\\\addlinespace
Epiphany-V & Adapteva & 1024 & 2048 * F & 117 & 4.5B & TBD & 16FF+ &
\\\addlinespace
\bottomrule
\addlinespace
\caption{Processor Comparisons. Nodes are programmable elements that can
execute independent programs, FLOPS are 64-bit floating point
operations, Area is expressed in mm\^{}2. Epiphany performance is
expressed in terms of Frequency (``F'').}
\end{longtable}

The correlation between silicon area processing efficiency and and
energy efficiency is well established. A processor with less active
silicon will generally have a higher energy efficiency. The table below
compares silicon efficiency and energy efficiency of modern processors.

\begin{longtable}[c]{@{}llll@{}}
\toprule\addlinespace
Chip & GFLOPS/mm\^{}2 & GFLOPS/W & W/mm\^{}2
\\\addlinespace
\midrule\endhead
P100 & 7.7 & 18.8 & 0.40
\\\addlinespace
KNL & 5.27 & 14.69 & 0.35
\\\addlinespace
Broadwell & 2.85 & 9.08 & 0.31
\\\addlinespace
Epiphany-V & 8.55 & TBD & TBD
\\\addlinespace
\bottomrule
\addlinespace
\caption{Normalized Double Precision Floating Point Peak Performance
Numbers. An arbitrary 500MHz operating frequency is used for
Epiphany-V.}
\end{longtable}

\begin{longtable}[c]{@{}lll@{}}
\toprule\addlinespace
Chip & Nodes/mm\^{}2 & MB RAM / mm\^{}2
\\\addlinespace
\midrule\endhead
P100 & 0.09 & 0.034
\\\addlinespace
KNL & 0.11 & 0.05
\\\addlinespace
Broadwell & 0.05 & 0.15
\\\addlinespace
Epiphany-V & 8.75 & 0.54
\\\addlinespace
\bottomrule
\addlinespace
\caption{64-bit Processors Metrics Normalized to Silicon Area}
\end{longtable}

\subsection{VIII. Conclusions \& Future
Work}\label{viii.-conclusions-future-work}

In this work we described the design of a 16nm parallel processor with
1024 64-bit RISC cores. The design was completed at 1/100th the cost of
the status quo and demonstrates an 80x advantage in processor density
and 3.6x-15.8x advantage in memory density compared to state of the art
processors.

Given the demonstrated order of magnitude silicon efficiency advantage,
Epiphany-V shows promise for the silicon limited Post-Moore era.

The next task will be to to fully characterize the Epiphany-V silicon
devices once devices return from the foundry. Future work will focus on
extending and customizing the Epiphany-V SOC platform for specific
target applications.

\subsection{Acknowledgment}\label{acknowledgment}

\begin{itemize}
\itemsep1pt\parskip0pt\parsep0pt
\item
  DARPA/MTO - For keeping chip research alive and well in the US
\item
  Ericsson - For being an outstanding partner
\item
  Parallella Backers - For taking a chance when others wouldn't
\item
  Ola Jeppsson - For being a software renaissance man
\item
  Roman Trogan - For always having my back (and for Epiphany 2-4)
\item
  Oleg Raikhman - For always having my back (and for Epiphany 2-4)
\item
  Shiri Jackman - For keeping the Adapteva office sane for 3 tough years
\item
  Graham Celine - For helping me get past the Kickstarter hurdles
\item
  Diana South - For keeping the mob at bay
\item
  David Richie - For showing what Epiphany can do in the right hands
\item
  Jeremy Bennett - For creating the first Epiphany SDK
\item
  Prof Jan Van Der Spiegel - For getting me hooked on electrical
  engineering
\item
  Prof Fay Ajzenberg - For showing kindness and courage
\item
  Prof Nader Engheta - For humble brilliance
\item
  Prof Ken Laker - For getting me my first job
\item
  Martin Izzard - For teaching me corporate professionalism
\item
  Fredy Lange - For igniting my passion for chip design
\item
  Yehuda Adelman - For being a role model
\item
  Henri Meirov - For teaching me the value of methodology
\item
  Kevin Leary - For invaluable strategy insight
\item
  Paul Kettle - For being an architecture sounding board
\item
  Ray Stata - For encouragement and mentorship
\item
  Mikael Taveniku - For embedded HPC insight
\item
  Jeff Milrod - For taking a chance on me
\item
  Blaise Aguire - For being a great friend
\item
  Ulla Olofsson (Mamma) - For always being there
\item
  Rolf Olofsson (Pappa) - For teaching me to always follow my passion
\item
  Sharon Olofsson - For everything
\end{itemize}

\subsection{References}\label{references}

\begin{itemize}
\item
  {[}1{]} Cadence Blog,
  \href{http://community.cadence.com/cadence_blogs_8/b/ii/archive/2009/09/24/are-soc-development-costs-significantly-underestimated}{Are
  SoC Development Costs Significantly Underestimated?}
\item
  {[}2{]} Ed Sperling,
  \href{http://semiengineering.com/how-much-will-that-chip-cost}{How
  Much Will That Chip Cost?}, Semi Engineering, March 27, 2014
\item
  {[}3{]} Rick Merritt,
  \href{http://www.eetimes.com/document.asp?doc_id=1261542\&page_number=2}{Nvidia
  calls for move to 450mm wafers}, EE Times, April 2012
\item
  {[}4{]} Ron Wilson,
  \href{http://www.eetimes.com/document.asp?doc_id=1177397}{Designers
  report on multi-million gate ASICs}, EE Times, June 2002
\item
  {[}5{]} A. Olofsson, Mesh network,
  \href{https://www.google.com/patents/US8531943}{U.S. Patent
  8,531,943}, Filed October 29 2009, Issued September 10, 2013,
\item
  {[}6{]} A. Olofsson, R. Trogan, and O. Raikhman,
  \href{http://www.adapteva.com/wp-content/uploads/2011/10/adapteva_hpec11.pdf}{A
  1024-core 70 GFLOP/W Floating Point Manycore Microprocessor}, 15th
  Annual Workshop on High Performance Embedded Computing, Sept 2011
\item
  {[}7{]} Adapteva,
  \href{http://www.adapteva.com/docs/e16g301_datasheet.pdf}{E16G301
  Datasheet}
\item
  {[}8{]} Adapteva,
  \href{https://www.kickstarter.com/projects/adapteva/parallella-a-supercomputer-for-everyone}{Parallella
  Kickstarter Project}
\item
  {[}9{]} Parallella Publication List,
  https://parallella.org/publications
\item
  {[}10{]} A. Olofsson, T. Nordström and Z. Ul-Abdin,
  \href{http://arxiv.org/pdf/1412.5538v1.pdf}{Kickstarting
  High-performance Energy-efficient Manycore Architectures with
  Epiphany} 2014 48th Asilomar Conference on Signals, Systems and
  Computers, Pacific Grove, CA, 2014, pp.~1719-1726
\item
  {[}11{]} A. Olofsson,
  \href{http://adapteva.com/docs/epiphany_arch_ref.pdf}{Epiphany
  Architecture Reference Manual}
\item
  {[}12{]} Alexandros Papadogiannakis, Spiros N. Agathos, Vassilios V.
  Dimakopoulo,
  \href{http://link.springer.com/chapter/10.1007/978-3-319-24595-9_15}{OpenMP
  4.0 Device Support in the OMPi Compiler}, OpenMP: Heterogenous
  Execution and Data Movements, Volume 9342 of the series Lecture Notes
  in Computer Science pp 202-216
\item
  {[}13{]} James Ross, David Richie, Song Park, Dale Shires,
  \href{http://www.sciencedirect.com/science/article/pii/S0141933116000375}{Parallel
  programming model for the Epiphany many-core co-processor using
  threaded MPI}, Microprocessors and Microsystems, Volume 43, Pages
  95-103, June 2016
\item
  {[}14{]} James Ross,David Richie,
  \href{http://arxiv.org/pdf/1608.03545v1.pdf}{An OpenSHMEM
  Implementation for the Adapteva Epiphany Coprocessor}, OpenSHMEM 2016,
  Third workshop on OpenSHMEM and Related Technologies
\item
  {[}15{]} David Richie, James Ross,
  \href{http://arxiv.org/pdf/1608.03549v1.pdf}{OpenCL + OpenSHMEM Hybrid
  Programming Model for the Adapteva Epiphany Architecture}, OpenSHMEM
  2016, Third workshop on OpenSHMEM and Related Technologies
\item
  {[}16{]} Magnus Lang, Kostis Sagonas,
  \href{http://www.erlang-factory.com/static/upload/media/143921284148554erlangonparallellaeuc2015.pdf}{Running
  Erlang on the Parallella}, Erlang User Conference 2015
\item
  {[}17{]} Jan-Willem Buurlage, Tom Bannink, Abe Wits,
  \href{https://arxiv.org/pdf/1608.07200v1.pdf}{Bulk-synchronous
  pseudo-streaming algorithms for many-core accelerators}, HLPP 2016:
  International Symposia on High-Level Parallel Programming and
  Applications
\item
  {[}18{]} Nick Brown, Epython Library,
  https://github.com/mesham/epython
\item
  {[}19{]} Andreas Olofsson, et al, PAL Library,
  https://github.com/parallella/pal
\item
  {[}20{]} Andreas Olofsson, et al, OH! Library,
  https://github.com/parallella/oh
\item
  {[}21{]} Andreas Olofsson,
  \href{http://www.gsaglobal.org/wp-content/uploads/2013/10/20120501_A_Lean_Fabless_Semiconductor_Business_Model.pdf}{A
  Lean Fabless Semiconductor Business Model}, 2012 GSA Global
\item
  {[}22{]} Clive Maxfield,
  \href{http://www.eetimes.com/electronicsblogs/other/4211089/From-RTL-to-GDSII-in-Just-Six-Weeks-}{From
  RTL to GDSII in Just Six Weeks}, EETimes Nov 2011
\item
  {[}23{]} NVIDIA,
  \href{https://devblogs.nvidia.com/parallelforall/inside-pascal}{Inside
  Pascal}
\item
  {[}24{]} Ian Cutress,
  \href{http://www.anandtech.com/show/9802/supercomputing-15-intels-knights-landing-xeon-phi-silicon-on-display}{SuperComputing
  15: Intel's Knights Landing Xeon Phi Silicon on Display}, November 19,
  2015
\item
  {[}25{]} Johan De Gelas,
  \href{http://www.anandtech.com/show/10401/intel-launches-4s-and-8s-broadwellex-xeons-e74800-v4-and-e78800-v4-families-up-to-24-cores}{The
  Intel Xeon E5 v4 Review: Testing Broadwell-EP With Demanding Server
  Workloads}, Anandtech, June 13 2016
\item
  {[}26{]} Brent Bohnenstiehl, Aaron Stillmaker, Jon Pimentel, Timothy
  Andreas, Bin Liu, Anh Tran, Emmanuel Adeagbo and Bevan Baas, KiloCore:
  A 32 nm 1000-Processor Array, IEEE HotChips Symposium on
  High-Perfomance Chips, August 2016
\end{itemize}

\end{document}